# Capturing the Variabilities of Distribution Network Net-Load via Available Flexibility of Microgrids

A. MAJZOOBI, A. KHODAEI
University of Denver
USA

S. BAHRAMIRAD
ComEd
USA

M. BOLLEN
Luleå Univ of Technology
Sweden

**SUMMARY**

Renewable energy has attracted significant attention over the last decade, conceivably due to its environmental benefits and the recent drops in the development and deployment cost of the technology. The increase in renewable generation, however, has resulted in new challenges in supply-load balancing, owing to its intermittent, non-predictable and volatile generation features. Several methods have been introduced to cope with negative impacts of the renewable generation deployment. In this paper, a novel method, i.e., the application of microgrids in capturing the variabilities of distributed renewable generation in distribution networks is proposed and investigated. Utilizing available flexibility of microgrids represents a local and viable solution which leads to lower investments from electric utilities for increasing their flexibility and providing more reserved power. It is investigated that how the system flexibility requirements can be integrated into the microgrid optimal scheduling model to enable microgrids in supporting the grid operators by offering flexibility services. Using the proposed flexibility constraints, intra-hour and inter-hour variabilities at the distribution feeder will be efficiently captured. Numerical simulations on a test distribution feeder, with one microgrid and several renewable-equipped consumers, show the effectiveness of the proposed model.

Alireza.Majzoobi@du.edu

# 1. INTRODUCTION

Renewable energy resource deployment has experienced a significant global growth in recent years, however, the renewable generation's inherent intermittency and volatility have caused new challenges in balancing electricity supply and demand, thus calling for new methods to diminish their negative impacts on power system stability and reliability [1]. There have been interesting discussions on how to address variability in voltage, beyond flicker, in distribution network [2]-[4]. The other significant challenge is how to maintain supply-demand balance. To address this challenge, grid operators traditionally rely on bulk power generation resources, such as fast ramping hydro and thermal units, that can be quickly dispatched and ramped up, or on distributed resources, such as demand response and electric vehicles, that in aggregate can provide considerable flexibility services. The large-scale solution, however, is constrained with the limited number and capacity of available resources, and is capital-intensive and time-consuming to be constructed. The small-scale solutions, on the other, are constrained with the requirement of a strong infrastructure, in terms of monitoring, and customers' willingness to participate in the flexibility offering, in terms of control. Energy storage, is another solution which is still rather expensive and cannot be extensively utilized in the power system.

Available flexibility in existing microgrids can potentially offer a solution for addressing renewable generation challenges and ensuring supply-load balance [5], [6]. A microgrid is defined as a group of interconnected loads and distributed energy resources (DERs) within clearly defined electrical boundaries that acts as a single controllable entity with respect to the grid and can connect and disconnect from the grid to enable it to operate in both grid-connected or island-mode [7]. The microgrid, as a novel distribution network architecture, consists of local and distributed generation resources, energy storage, and loads. These resources are controlled and operated by a microgrid master controller, based on economy and reliability considerations, thus the microgrid can be seen as a relatively large and controllable demand from a grid operator's perspective. The microgrid has the capability to disconnect from the utility grid and operate in the islanded mode, in order to isolate the microgrid from any disturbances in the upstream network and minimizing the amount of local load curtailment [8]. During the grid-connected mode, however, microgrid freely exchanges power with the utility grid which can be in the form of power import or export. If this power exchange is efficiently controlled, by adding proper constraints to the microgrid optimal scheduling framework, the microgrid can be used to capture the generation variability of distributed renewable energy resources in the distribution grid. These flexibility constraints, that enable the microgrid to support the utility grid in terms of flexibility services, are investigated and developed in the following sections of this paper.

The proposed solution in this paper can be of significant value particularly in future power grids where the penetration of microgrids is high. A global trend can be seen in microgrid deployments, where the microgrids revenue is anticipated to reach $19.9 billion by 2020 [9]. More than 1,500 microgrid projects, with the capacity of 15,600 MW, have been reported until April 2016 and 1,565 MW new microgrid projects have been introduced in 39 countries as new projects [10]. These figures clearly show the growing interest in microgrids and further signify microgrids as core components of future power grids. The rest part of the paper is organized as follows: The outline of the proposed constraints along with mathematical modelling are presented in Section 2, numerical simulations on a test feeder to show the effectiveness of proposed model are provided in Section 3, and the paper is concluded in Section 4.

## 2. MODEL OUTLINE AND FORMULATION

Figure 1 depicts the impact of the microgrid on a distribution feeder consisting of a microgrid and distributed prosumers/consumers. Prosumers are consumers that are equipped with distributed generation resources, primarily renewable energy. This figure compares the impact of the microgrid on the distribution feeder net load, from the utility grid's perspective, with and without consideration of flexibility constraints in the microgrid scheduling. In the top figure, the microgrid is scheduled to obtain the least-cost operation, i.e., local resources are scheduled in a way that the microgrid operation cost (consisting of local generation cost and cost of energy exchange with the utility grid) is minimized. In the bottom figure, however, the flexibility constraints are considered in the microgrid optimal scheduling framework, i.e., local resources are scheduled in a way that the flexibility requirement by the utility grid are met with the support of the microgrid. In the former case, the microgrid not only does not mitigate the net load variabilities in the distribution feeder, but also exacerbates the variabilities. In the latter case, however, the microgrid captures the existing variabilities where the distribution feeder net load profile, seen by the utility grid, is considerably smoother and variabilities are confined to desired amount.

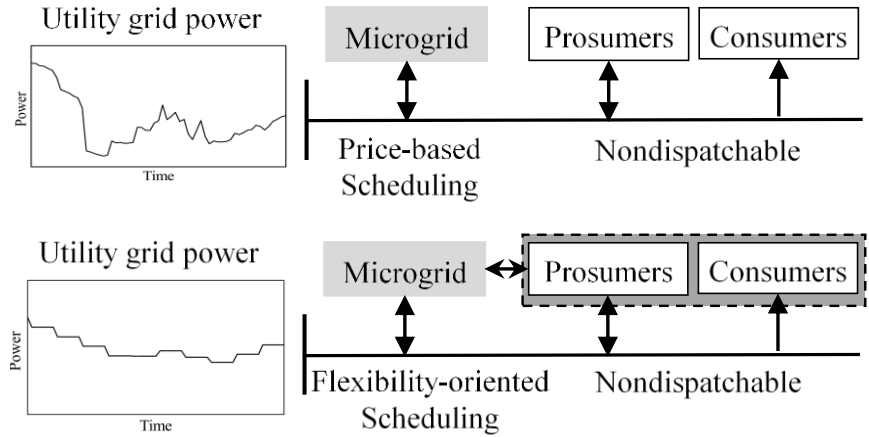

Figure 1. Impact of microgrid scheduling on a distribution feeder; top: price-based scheduling, bottom: Flexibility-oriented scheduiling.

The amount of the power that is to be provided by the utility grid, which is also the distribution feeder net load, can be calculated based on the microgrid net load and prosumers/consumers net load as in (1):

$$P_{tks}^{u} = P_{tks}^{M} + \sum_{j \in N} P_{jtks}^{c} \qquad \forall t, \forall k, \forall s, \qquad (1)$$

where, $P^u$ represents the utility grid power, $P^M$ represents the microgrid exchanged power with the utility grid (microgrid net load), and $P^c$ shows the distributed prosumers/consumers net load. $t$ and $k$ denote indices for inter-hour and intra-hour time periods, respectively, and $s$ is the index for various scenarios of microgrid islanding which is considered for considering microgrid operation in the islanded mode. Different values can be considered for $k$ as the intra-hour time period to enable modelling different time resolutions, such as 1-minute based, 5-minute based, 10-minute based, etc. The grid operator is required to limit the intra-hour and inter-hour variabilities in distribution feeders, to minimize the necessary flexibility resource investments, as proposed in (2) and (3):

$$\left| P_{tks}^u - P_{t(k-1)s}^u \right| \leq \Delta_1 \qquad \forall t, \forall s, k \neq 1, \qquad (2)$$

$$\left| P_{t1s}^u - P_{(t-1)Ks}^u \right| \leq \Delta_2 \qquad \forall t, \forall s, \qquad (3)$$

where $\Delta_1$ and $\Delta_2$, respectively the utility grid's intra-hour and inter-hour desired ramping, are selected by the grid operator based on the day-ahead net load forecasts and desired grid flexibility during each time interval. If the required grid flexibility is larger than the available flexibility, which is calculated using various methods [11] and [12], the grid operator can utilize the available flexibility of distributed resources, such as microgrids, to compensate for the flexibility. Thus, $\Delta_1$ and $\Delta_2$ are defined by comparing the required and available grid flexibility. Moreover, the grid operator can obtain these limits through cost-benefit analysis. To this end, the expense for increasing the utility grid flexibility will be compared with the payments to the participating microgrids. These two constraints can be converted into constraints on the microgrid net load by substituting the value of $P^u$ from (1). Consequently, (4) and (5) will be obtained:

$$-\Delta_1 - (\sum_j P_{jtk}^c - \sum_j P_{jt(k-1)}^c) \leq P_{tks}^M - P_{t(k-1)s}^M \leq \Delta_1 - (\sum_j P_{jtk}^c - \sum_j P_{jt(k-1)}^c) \qquad \forall t, \forall s, k \neq 1, \qquad (4)$$

$$-\Delta_2 - (\sum_j P_{jt1}^c - \sum_j P_{j(t-1)K}^c) \leq P_{t1s}^M - P_{(t-1)Ks}^M \leq \Delta_2 - (\sum_j P_{jt1}^c - \sum_j P_{j(t-1)K}^c) \qquad \forall t, \forall s. \qquad (5)$$

These two constraints can be added to the microgrid optimal scheduling problem to model the flexibility constraints and therefore utilize the microgrid for grid support. It should also be noted that the utility grid flexibility constraints, which are obtained by utility grid ($\Delta_1$ and $\Delta_2$), are fixed while as the boundaries of constraints (4) and (5) show, the limits on microgrid net load are variable. This is due to aggregated net load of all consumers/prosumers in distribution feeder which is highly variable. In addition, changing intra-hour period from 1 to 60 enables having different resulotions in the model from one minute to one hour. Therefore, the grid operator can reach to a desired profile of net load in distribution network with managing both intra-hour and inter-hour flexibility limits.

The complete formulation of the microgrid optimal scheduling, along with prevailing technical constraints, can be found in the literature [13].

## 3. SIMULATION RESULTS

A microgrid with four dispatchable units, two nondispatchable units (solar and wind), one energy storage, and five adjustable loads is considered for simulations. The characteristics of microgrid's components and the hourly market price are available in [13]. This microgrid is connected to a distribution feeder with several prosumers and consumers. The aggregated net load of these prosumers/consumers is shown in Figure 2.

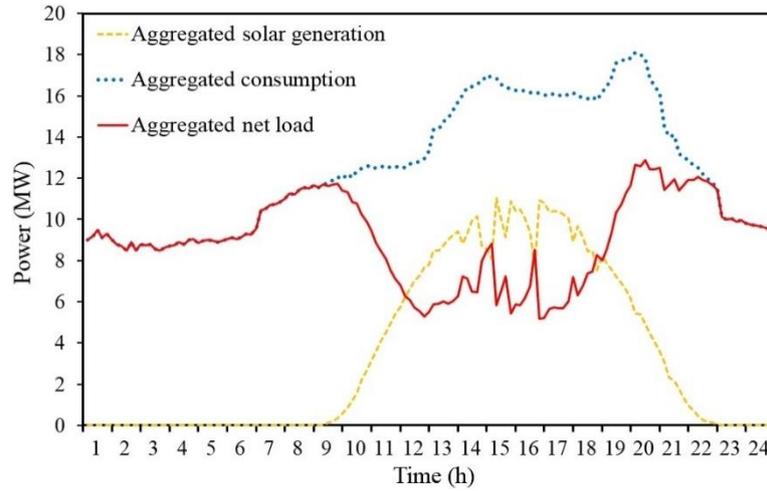

Figure 2. Aggregated prosumers/consumers generation, consumption, and net load.

The microgrid optimal scheduling problem is solved using price-based scheduling (no flexibility constraints are considered) and flexibility-oriented scheduling (the proposed flexibility constraints are considered). Figure 3 shows the distribution feeder net load, seen by the utility grid, in these two studied cases. From this figure it can be easily deduced that under the price-based scheduling, the microgrid exacerbates the variabilities of distribution feeder net load due to the fact that in this condition the microgrid controller schedules its local generation and loads to achieve the least operation cost without consideration of required flexibility of utility grid. However, under flexibility-oriented scheduling, as Figure 3 shows, a smoother net load profile is seen by the utility grid and the variabilities of distribution feeder net load are restricted to the desired flexibility limits. A 0 MW/10-min is considered as the intra-hour variability limit ($\Delta_1$), therefore all intra-hour variabilities of distribution feeder net load are captured by the microgrid. The microgrid net load is depicted in Figure 4 to show how microgrid revises its scheduling, and accordingly its utility grid power exchange, to capture distribution feeder variabilities. As this figure shows, there are considerable variabilities in the microgrid net load under flexibility-oriented scheduling as a result of capturing intra-hour variabilities of distribution feeder net load. In price-based scheduling, however, the microgrid imports the maximum possible power from the utility grid when electricity prices are low, whilst reduces the amount of import and even exports excess power to the utility grid in high electricity price to achieve its primary objective, i.e. the minimum operation cost.

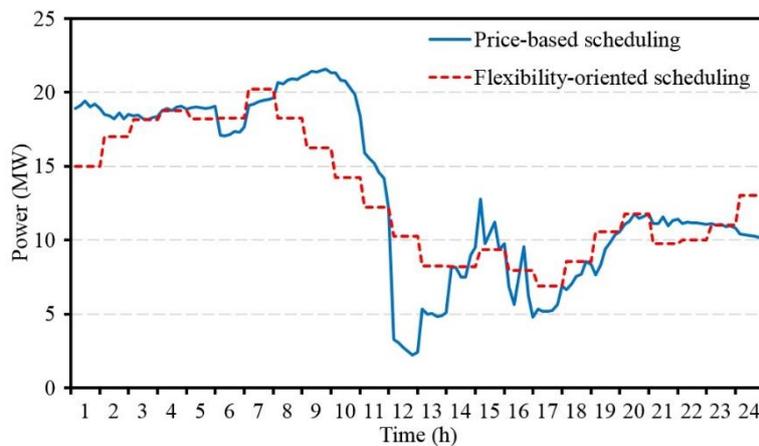

Figure 3. Distribution feeder net load in two conditions: price-based scheduling and Flexibility-oriented scheduling.

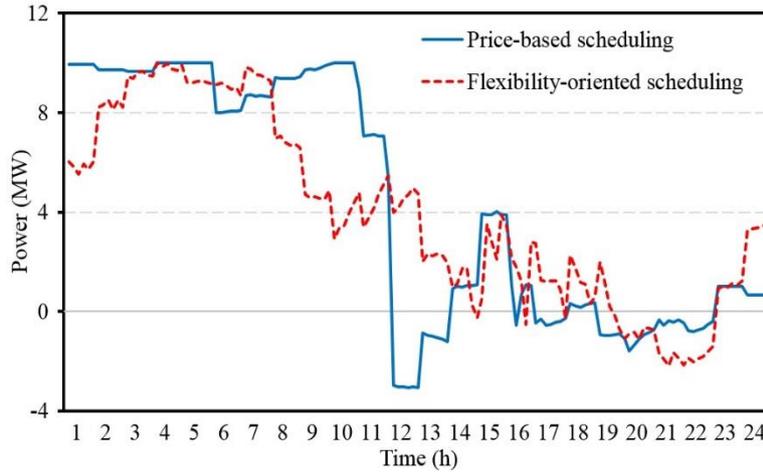

Figure 4. Microgrid net load in two conditions: price-based scheduling and Flexibility-oriented scheduling.

The results of the two cases demonstrate the effectiveness of the proposed model. However, it should be mentioned that the microgrid operation cost will be increased owing to adding flexibility constraints to the problem. The microgrid operation cost considering the flexibility constraints is calculated as $13,329 which is significantly increased, around 13.5%, compared to that of the price-based scheduling. This increase in the microgrid operation cost, which is imposed to the microgrid due to considering flexibility constraints, should be paid to the microgrid as an incentive for contribution in grid support and reducing distribution feeder net load variabilities. This cost increase may introduce a good argument to solve the problem centrally, by the grid operator, and further determine the incurred system costs. This analysis, however, requires further development of appropriate grid operation and planning schemes which will be considered as follow up works.

## 4. CONCLUSIONS

Microgrids, as local and highly flexible resources, can be used to capture the distribution grid net load variabilities caused primarily by distributed renewable energy resources. This paper investigated this capability and proposed novel constraints that can be added to the microgrid optimal scheduling problem and enable grid support. The proposed constraints, which were developed based on inter-hour and intra-hour time indices, could potentially capture high resolution (minute-based) load variabilities, thus would be of significant value in managing variable renewable generation using microgrids. Simulation results on a test feeder with one microgrid revealed the effectiveness of the proposed model while showing that this benefit comes at the expense of an increase in the microgrid operation cost.